\documentclass[letter,traditabstract,longauth]{aa} % for the long lists of affiliations
\usepackage{graphicx} %%%%%%%%%%%%%%%%%%%%%%%%%%%%%%%%%%%%%%%% 
\begin{document} %
\def\simlt{\mathrel{\rlap{\lower 3pt\hbox{$\sim$}}\raise 2.0pt\hbox{$<$}}}
\def\simgt{\mathrel{\rlap{\lower 3pt\hbox{$\sim$}} \raise
2.0pt\hbox{$>$}}}

\title{\textit{Herschel}-ATLAS: blazars in the SDP field\thanks{\textit{Herschel} is an ESA space observatory with science instruments provided by European-led Principal Investigator consortia and with important participation from NASA.}}

   \subtitle{}
   \author{J. Gonz\'alez-Nuevo
          \inst{1}\fnmsep\thanks{e-mail: gnuevo@sissa.it}
          \and
	G. De Zotti\inst{2,1}
    \and
    P. Andreani\inst{3,4}
    \and
    E.J. Barton\inst{5}
    \and
    F. Bertoldi\inst{6}
    \and
    M. Birkinshaw\inst{7}
    \and
    L. Bonavera\inst{1,8}
    \and
    S. Buttiglione\inst{2}
    \and
    J. Cooke\inst{5}
    \and
    A. Cooray\inst{5}
    \and
    G. Danese\inst{1}
    \and
	L. Dunne\inst{9}
	\and          
	S. Eales\inst{10}
	\and
	L. Fan\inst{1}
    \and
    M.J. Jarvis\inst{11}
    \and
    H-R. Kl\"ockner\inst{12}
    \and
	E. Hatziminaoglou\inst{3}
    \and
    D. Herranz\inst{13}
    \and
    D.H. Hughes\inst{14}
    \and
    A. Lapi\inst{15,1}
    \and
    A. Lawrence\inst{16}
    \and
    L. Leeuw\inst{17}
    \and
    M. Lopez-Caniego\inst{13}
    \and
    M. Massardi\inst{2}
    \and
    T. Mauch\inst{12}
    \and
    M.J. Micha{\l}owski\inst{16}
    \and
	M. Negrello\inst{18}
	\and          
    S. Rawlings\inst{12}
    \and
    G. Rodighiero\inst{19}
    \and
    S. Samui\inst{1}
    \and
    S. Serjeant\inst{18}
    \and
    J.D. Vieira\inst{20}
    \and
    G. White\inst{18,21}
          \and
   	A. Amblard\inst{5}
	\and
	R. Auld\inst{10}
	\and
	M. Baes\inst{22}
	\and
	D.G. Bonfield\inst{11}
	\and
	D. Burgarella\inst{23}
	\and
	A. Cava\inst{24,25}
	\and
	D.L. Clements\inst{26}
	\and
	A. Dariush\inst{10}
	\and
	S. Dye\inst{10}
	\and
	D. Frayer\inst{27}
	\and
	J. Fritz\inst{22}
	\and
	E. Ibar\inst{28}
	\and
	R.J. Ivison\inst{28}
	\and
	G. Lagache\inst{29}
	\and
	S. Maddox\inst{9}
	\and
	E. Pascale\inst{10}
	\and
	M. Pohlen\inst{10}
	\and
	E. Rigby\inst{9}
	\and
	B. Sibthorpe\inst{28}
	\and
	D.J.B. Smith\inst{9}
	\and
	P. Temi\inst{17}
	\and
	M. Thompson\inst{11}
	\and
	I. Valtchanov\inst{30}
	\and
	A. Verma\inst{12}
}	
   \institute{
	%1:
	SISSA, Via Beirut 2--4, I-34014 Trieste, Italy
	\and
	%2:
	INAF-Osservatorio Astronomico di Padova, Vicolo dell'Osservatorio 5, I-35122 Padova, Italy
	\and
    %3:
	ESO, Karl-Schwarzschild-Str.2, D-85748, Garching, Germany
    \and
	%4:
	INAF, Osservatorio Astronomico di Trieste, Via Tiepolo 11, I-34143, Trieste, Italy
    \and
	%5:
	Department of Physics \& Astronomy, University of California, Irvine, CA 92697, USA
  	\and
  	%6:
  	Argelander Institute for Astronomy, University of Bonn, Auf dem H\"ugel 71, 53121 Bonn, Germany
    \and
    %7:
    Department of Physics, University of Bristol, Tyndall Avenue, Bristol BS8 1TL, U.K.
	\and
	%8:
	Australia Telescope National Facility, CSIRO, PO Box 76, Epping, NSW 1710, Australia
	\and
    %9:
    School of Physics and Astronomy, University of Nottingham, University Park, Nottingham NG7 2RD, UK
    \and
    %10:
    School of Physics and Astronomy, Cardiff University, The Parade, Cardiff, CF24 3AA, UK
    \and
    %11:
    Centre for Astrophysics, Science \& Technology Research Institute, University of Hertfordshire, Hatfield, AL10 9AB, UK
    \and
    %12:
    Astrophysics, Department of Physics, University of Oxford, Keble Road, Oxford OX1 3RH, U.K.
    \and
    %13:
    Instituto de F\'\i{sica} de Cantabria (CSIC-UC), Avda. los Castros s/n, 39005 Santander, Spain
    \and
    %14:
    Instituto Nacional de Astrof\'{i}sica, \'{O}ptica y Electr\'{o}nica (INAOE),
Luis Enrique Erro No.1, Tonantzintla, Puebla, C.P. 72840, Mexico
	\and
	%15:
	Dipartimento di Fisica, Universita Tor Vergata, Via Ricerca Scientifica 1, 00133 Roma, Italy
	\and
	%16:
    Institute for Astronomy, University of Edinburgh, Royal Observatory, Blackford Hill, Edinburgh EH9 3HJ, U.K.		
    \and
    %17:
    Astrophysics Branch, NASA Ames Research Center, Mail Stop 245-6, Moffett Field, CA 94035, USA
	\and
	%18:
    Department of Physics \& Astronomy, The Open University, Milton Keynes MK7 6BJ, U.K.	    \and
    %19:
	Department of Astronomy, University of Padova, Vicolo dellOsservatorio 3, I-35122 Padova, Italy
	\and
	%20:
	Division of Physics, Mathematics \& Astronomy, California Institute of Technology, Mail Code 59-33, Pasadena, CA 91125
	\and
	%21:
	Space Science Department, Rutherford Appleton Laboratory, Chilton, UK
	\and
	%22:
	Sterrenkundig Observatorium, Universiteit Gent, Krijgslaan 281 S9, B-9000 Gent, Belgium		
	\and
	%23:
	Laboratoire d'Astrophysique de Marseille, UMR6110 CNRS, 38 rue F. Joliot-Curie, F-13388 Marseille France
	\and
	%24:
	Instituto de Astrof{\'\i}sica de Canarias (IAC), E-38200 La Laguna, Tenerife, Spain
	\and
	%25:
	Departamento de Astrof{\'\i}sica, Universidad de La Laguna (ULL), E-38205 La Laguna, Tenerife, Spain
	\and
	%26:
	Astrophysics Group, Imperial College, Blackett Laboratory, Prince Consort Road, London SW7 2AZ, UK
	\and
	%27:
	National Radio Astronomy Observatory,  PO Box 2, Green Bank, WV  24944, USA
	\and
	%28:
	UK Astronomy Technology Center, Royal Observatory Edinburgh, Edinburgh, EH9 3HJ, UK
	\and
	%29:
	Institut d'Astrophysique Spatiale (IAS), Batiment 121, F-91405 Orsay, France; and Universite Paris-Sud 11 and CNRS (UMR 8617), France
	\and
	%30:
	Herschel Science Centre, ESAC, ESA, PO Box 78, Villanueva de la Ca\~nada, 28691 Madrid, Spain
		             }

   \date{Received March 31, 2010; Accepted May 5, 2010}

% \abstract{}{}{}{}{} % 5 {} token are mandatory

 \abstract{
  % context heading (optional)
  % {} leave it empty if necessary
   {}
  % aims heading (mandatory)
{To investigate the poorly constrained sub-mm counts and spectral properties of blazars we searched for these in the \textit{Herschel}-ATLAS (H-ATLAS) science demostration phase (SDP) survey catalog.}
  % methods heading (mandatory)
{}
 % results heading (mandatory)
{We cross-matched 500$\mu$m sources brighter than 50 mJy with the FIRST radio catalogue. We found two blazars, both previously known. Our study is among the first blind blazar searches at sub-mm wavelengths, i.e., in the spectral regime where little is still known about the blazar SEDs, but where the synchrotron peak of the most luminous blazars is expected to occur. Our early results are consistent with educated extrapolations of lower frequency counts and question indications of substantial spectral curvature downwards and of spectral upturns at mm wavelengths.  One of the two blazars is identified with a  Fermi/LAT $\gamma$-ray source and a WMAP source. The physical parameters of the two blazars are briefly discussed. }
% conclusions heading (optional), leave it empty if necessary
{These observations demonstrate that the H-ATLAS survey will provide key information about the physics of blazars and their contribution to sub-mm counts.}}


\maketitle

\begin{table*}
\caption{The two H-ATLAS blazars found by cross-matching with the FIRST catalog. Coordinates (RA and Dec) refer to the H-ATLAS source.  $\Delta$ (arcsec) is the angular separation between FIRST and \textit{Herschel} positions. The quoted errors in \textit{Herschel} fluxes are statistical only. Calibration errors are estimated to be $\sim$15\% for SPIRE (Swinyard et al. 2010), $\sim$10\% at $100\,\mu$m, and  $\sim$20\% at $160\,\mu$m. Upper limits are at the 3$\sigma$ level.}\label{tab:candidates}
\begin{center}
\begin{tabular}{lccccccc}
\multicolumn{1}{c}{H-ATLAS} & \multicolumn{1}{c}{$\Delta$} & \multicolumn{1}{c}{$S_{1.4{\rm GHz}}$} & \multicolumn{1}{c}{$S_{500\mu{\rm m}}$} & \multicolumn{1}{c}{$S_{350\mu{\rm m}}$} & \multicolumn{1}{c}{$S_{250\mu{\rm m}}$} & \multicolumn{1}{c}{$S_{170\mu{\rm m}}$} & \multicolumn{1}{c}{$S_{100\mu{\rm m}}$} \\
\multicolumn{1}{c}{name} & \multicolumn{1}{c}{arcsec} & \multicolumn{1}{c}{mJy} & \multicolumn{1}{c}{mJy} & \multicolumn{1}{c}{mJy} & \multicolumn{1}{c}{mJy} & \multicolumn{1}{c}{mJy} & \multicolumn{1}{c}{mJy} \\ \hline
J090910.1+012135$^a$ & 0.3 & 571.7 & $265.8 \pm 8.9$& $193.8\pm 7.2$ & $159.6\pm 6.4$ & $105.4\pm 26.5$ & 137.3$\pm 26.5$ \\
J090940.3+015957$^b$ & 10.3 & 317.5 &  $68.2\pm 9.1$ &  $55.0\pm 7.2$ &  $35.5\pm 6.4$ & $<76.2$ & $<79.5$ \\ \hline
\multicolumn{8}{l}{$^a$ [HB89] 0906+015,\ \ $^b$PKS 0907+022}
\end{tabular}
\end{center}
\end{table*}

\section{Introduction} Blazars, comprising BL Lac objects and flat-spectrum radio quasars, are a class of active galactic nuclei characterized by remarkable properties: high luminosities at the two extreme ends of the electromagnetic spectrum (i.e., in the radio and $\gamma$-ray bands), rapid variability, apparent superluminal jet speeds, flat or inverted radio spectrum, and high polarization in the optical waveband. They are interpreted in terms of relativistic effects due to jets propagating at velocities  close to the speed of light almost along the line-of-sight. Their spectral energy distribution (SED) can be accounted for as a combination of synchrotron emission, peaking at a frequency $\nu_{p}^s$ that varies from $\sim 10^{12}\,$Hz to $\simgt 10^{19}\,$Hz (Nieppola et al. 2006), and inverse Compton scattering, assumed to be responsible for the $\gamma$-ray peak.

Padovani \& Giommi (1995) classified BL Lac objects into 3 subclasses (low, intermediate, and high frequency synchrotron peaked BL Lacs) depending on the value of $\nu_{p}^s$. Abdo et al. (2009) extended the classification to all blazars, subdivided into low (LSP; $\nu_{p}^s<10^{14}\,$Hz), intermediate (ISP; $10^{14}< \nu_{p}^s < 10^{15}\,$Hz), and high (HSP; $\nu_{p}^s>10^{15}\,$Hz) synchrotron peaked blazars. For many years, the blazar selection has relied mostly on either radio or X-ray surveys.  Radio surveys preferentially select LSPs, while X-ray surveys favor the HSPs. The resulting global distribution of $\nu_{p}^s$ is thus bimodal, although objects peaking at intermediate frequencies have also been found. The high sensitivity and nearly uniform sky coverage of the Fermi Gamma-ray Space Telescope made it a powerful tool for providing large blazar samples (Abdo et al. 2010).

In the present paper, we report on the blazars selected at sub-mm wavelengths in the $16\,\hbox{deg}^2$ \textit{Herschel}-ATLAS (H-ATLAS; Eales et al. 2010) science demonstration phase (SDP) field. The \textit{Herschel} data cover the poorly explored frequency range close to the synchrotron peak of the most luminous LSPs (Fossati et al. 1998; Padovani et al. 2006); $\nu_p^s$ carries crucial information about key physical parameters, such as the Lorentz factor $\gamma_p$ of emitting electrons,  the Doppler factor $\delta$, and the magnetic field strength $B$ ($\nu_p^s\propto \gamma_p^2\, \delta\, B$).

\begin{figure}
 \centering
 \includegraphics[scale=0.45]{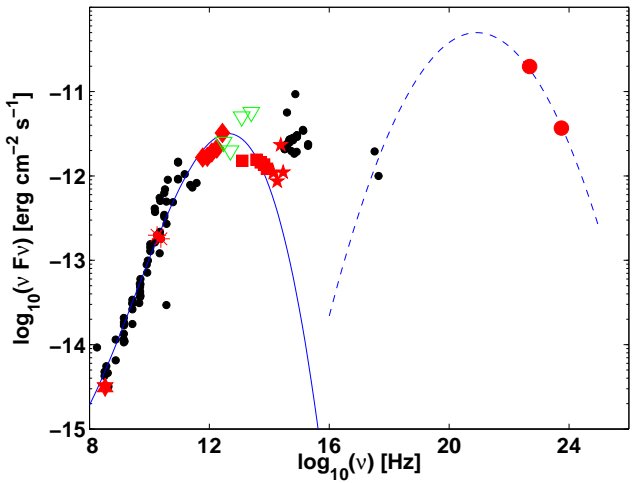}
   \caption{SED of [HB89] 0906+015 (non-simultaneous data), in the observer's frame. Black dots are data taken from the NASA/IPAC Extragalactic Database (NED), red diamonds refer to H-ATLAS data, red circles are the Fermi/LAT data, red asterisks are the new ATCA data, red squares are the new \textit{Spitzer} data, red stars are UKIDSS data, red hexagrams are the GMRT data, and green triangles are $3\sigma$ upper limits derived from IRAS maps. The blue solid/dashed lines  represent the best fit to the synchrotron/inverse Compton part of the SED, omitting the optical/UV fluxes, which are probably dominated by the direct emission from the accretion disk (see text). }
              \label{fig:sed1}%
    \end{figure}

Evolutionary models (De Zotti et al. 2005) predict $\simeq 0.15\,\hbox{blazars\ deg}^{-2}$ brighter than $50\,$mJy at $500\,\mu$m, the approximate $5\sigma$ detection limit of the H-ATLAS survey (Rigby et al. 2010, in preparation), so that the full survey may be expected to yield a sample of $\simeq 80$ blazars selected, for the first time, at sub-mm wavelengths. This prediction assumes a flat radio spectral index $\alpha$ ($\alpha\simeq 0.1$; $S_\nu \propto \nu^{-\alpha}$) followed by a parabolic decline close to and above a synchrotron peak frequency that increases with decreasing radio luminosity (Fossati et al. 2008).  However, spectral measurements have provided evidence of a substantial spectral curvature downwards from cm to mm wavelengths (Waldram et al. 2007; Sadler et al. 2008). Vieira et al. (2009) find that their highest significance 150 GHz selected radio sources are consistent with flat spectral behavior (or $\alpha \simeq 0.1$) between 5 GHz and 150 GHz but a possible steepening between 150 and 220 GHz. Voss et al. (2006) reported an excess (based however on the detection of only three blazars) above extrapolations from lower frequency counts of the surface density at $S\ge 10\,$mJy at 1.2 mm, suggesting that at least a fraction of blazars have inverted (i.e. rising with frequency) spectra at mm wavelengths, perhaps due to variability. The H-ATLAS data will also shed light on this issue.

In this paper, we describe (Sect. \ref{sec:cand}) the identification of blazars among sources detected in the H-ATLAS SDP field and discuss (Sect. \ref{sec:SED})  the properties of the 2 blazars found. In Sect. \ref{sec:concl} we summarize our main conclusions.

%__________________________________________________________________

\section{Blazar candidates}\label{sec:cand}

Observations were performed with the ESA \textit{Herschel} Space Observatory (Pilbratt et al. 2010). The $4\,\hbox{deg}\times 4\,\hbox{deg}$ H-ATLAS SDP field, centered on ($09^h 05^m 30^s$, $+00^\circ 30'00''$), was observed with the Spectral and Photometric Imaging Receiver (SPIRE; Griffin et al. 2010) at 250, 350, and $500\,\mu$m and with the Photodetector Array Camera and Spectrometer (PACS; Poglitsch et al. 2010) at 100 and $160\,\mu$m. Details of the SPIRE and PACS mapmaking are given by Pascale et al. (2010, in preparation) and Ibar et al. (2010, in preparation), respectively. Sources were extracted from the $250\,\mu$m map (see Rigby et al. 2010, in preparation). For each source, 350 and $500\,\mu$m flux densities were estimated by extracting flux from the appropriate noise-weighted beam-convolved map at the source position determined in the $250\,\mu$m map.  Fluxes at 100 and $160\,\mu$m were then assigned by matching to $\ge 3\,\sigma$ PACS sources within a positional tolerance of $10''$.

The selection of blazar candidates among H-ATLAS sources brighter than 50 mJy was made by identifying FIRST counterparts within a search radius of $15''$, the $\sigma_b \simeq \hbox{FWHM}/2.35$ of a Gaussian model of the 500$\mu$m beam. The probability that a true counterpart has an apparent positional offset $\ge \Delta$ is
\begin{equation}\label{eq:prob}
p(>\Delta) = \exp[-0.5(\Delta/\sigma)^2],
\end{equation}
where $\sigma$ is the global positional uncertainty, taking into account both H-ATLAS and FIRST positional errors. Sources detected with signal-to-noise ratio (S/N) greater than 10 at $250\,\mu$m, such as the blazar [HB89] 0906+015 (J090910.1+012135; see Table~\ref{tab:candidates}), have a $\sigma_{\rm H-ATLAS}\simeq 1.4''$. For PKS 0907+022 (J090940.3+015957), which has data of lower S/N at $250\,\mu$m, $\sigma_{\rm H-ATLAS}$ is $\simeq 4.9''$.
The FIRST positions are accurate to better than 1 arcsec at the survey limit and to better than 0.5 arcsec for sources brighter than 5 mJy at 1.4 GHz (White et al. 1997). With our choice of the search radius, the probabilities of having missed true associations at separations $\Delta > 15''$ are $\simeq 8\times 10^{-23}$ and $\simeq 10^{-2}$, respectively, for the two values of $\sigma_{\rm H-ATLAS}$ and $\sigma_{\rm FIRST}=0.5''$. We note that the two values of $\sigma_{\rm H-ATLAS}$ are representative of sources with data of high and low S/N at $250\,\mu$m.

Since the distribution of FIRST sources is to a large degree uniform, the probability that one of them lies by chance within the angular radius $\Delta$  from a given \textit{Herschel} source is
\begin{equation}\label{eq:nc}
n_c = \pi \Delta^2 n_{\rm 1.4GHz}(>S),
\end{equation}
where $S$ is the flux of the sources we match to.
The FIRST catalog is complete down to $0.75$ mJy/beam. The corresponding surface density is $n_{\rm 1.4GHz}(>0.75\,\hbox{mJy}) \simeq 89\,\hbox{deg}^{-2}$, implying that the maximum probability of a chance association within 15 arcsec is $5\times 10^{-3}$. Since there are 185 sources with $S_{500\mu{\rm m}}>50\,$mJy in the H-ATLAS SD field, the expected number of chance associations is $\simeq 0.9$. Simulations made by offsetting the radio positions at random and re-running the cross-matching yielded $n_c=0.9(+2.4,-0.8)$ (68\% confidence errors), consistent with expectations for Poisson statistics.

The cross-matching yielded 19 matches. Of them, 10 have far-IR/sub-mm colours typical of low-$z$ dusty galaxies. The far-IR/sub-mm SEDs of 4 others are consistent with those of dusty galaxies at $z\simeq 1$--2 and their radio emission obeys the radio-FIR correlation (Yun et al. 2001). The other three sources are more than  $4\sigma$  ($\sigma$ being the combined rms positional uncertainty) away from their nearest FIRST source, and are therefore very unlikely associations. We note that having 3 spurious associations within our search radius is fully consistent with the results of our simulations. All of these 17 sources were discarded. We note that we aim to select only sources with non-thermal emission brighter than 50 mJy at $500\,\mu$m, so we are not interested here in the weaker blazar nuclei that may be hiding in dusty galaxies. The FIRST and \textit{Herschel} photometric data for the 2 remaining sources are shown in Table~\ref{tab:candidates}.

Both H-ATLAS J090910.1+012135 ([HB89] 0906+015) and H-ATLAS J090940.3+015957 (PKS 0907+022) are known blazars. The former is a flat-spectrum quasar with strong broad emission lines  at a measured redshift of 1.018 (Falomo et al. 1994). The latter is a BL Lac with an estimated photometric redshift of 1.575 (Richard et al. 2004); a Keck spectrum obtained by E. Barton and J. Cooke does not show any clearly identifiable line. The surface densities of 1.4 GHz sources brighter than these sources are $\simeq 0.1\,\hbox{deg}^{-2}$ and $\simeq 1\,\hbox{deg}^{-2}$, respectively.  According to Eq.~(\ref{eq:nc}), the probability that they fall by chance within their angular distance, $\Delta$, from the H-ATLAS source is $\simeq 2\times 10^{-9}$ and $\simeq 2\times 10^{-5}$, respectively, while, after Eq.~(\ref{eq:prob}), the probability that the true counterparts to the H-ATLAS sources have an apparent angular separation as large as the measured one, or larger, is 99\% and 12\%, respectively.

\begin{figure}
 \centering
 \includegraphics[scale=0.45]{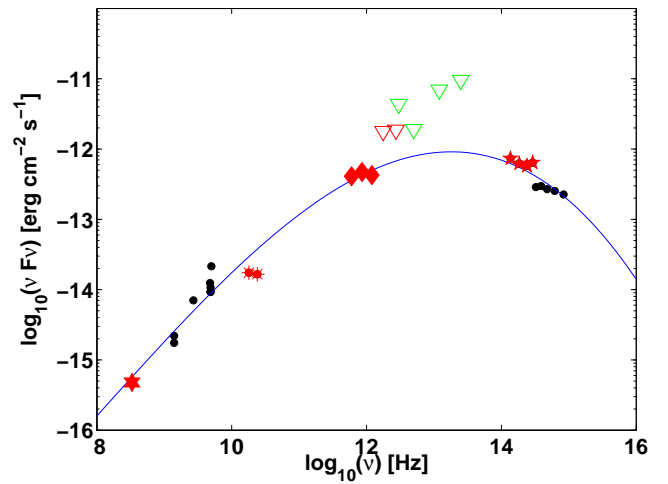}
   \caption{SED of PKS 0907+022 (non simultaneous data) in the observer's frame. Black dots: data from the NED; red diamonds and triangles: H-ATLAS flux measurements and $3\sigma$ upper-limits, respectively; red asterisks: ATCA measurements; red five-pointed starts: UKIDSS data; red six-pointed starts: GMRT data; green triangles: IRAS $3\sigma$ upper limits. The blue solid line is the best-fit relation for the synchrotron part of the SED. }
              \label{fig:sed2}%
    \end{figure}

\begin{table} \caption{GMRT, ATCA, UKIDSS, and \textit{Spitzer} flux density measurements and IRAS $3\sigma$ upper limits for the two blazars. }\label{tab:HA_fluxes}
\begin{center}
\begin{tabular}{lcc}
\hline
Flux (mJy) & [HB89] 0906+015 & PKS 0907+022\\
\hline
$S_{332\rm MHz}$ & 941 & 144\\
$S_{18\rm GHz}$ & $1101\pm 55$ & $97\pm 5$\\
$S_{24\rm GHz}$ & $742\pm 37$ & $69\pm 4$\\
$S_{1.03\mu\rm m}$ & $0.38$ & 0.22\\
$S_{1.25\mu\rm m}$ & $0.97$ & 0.24\\
$S_{1.63\mu\rm m}$ & $0.47$ & 0.34\\
$S_{2.20\mu\rm m}$ & $0.80$ & 0.54\\
$S_{3.6\mu\rm m}$ & $1.46\pm 0.09$ & --\\
$S_{4.5\mu\rm m}$ & $2.05\pm 0.10$ & --\\
$S_{5.8\mu\rm m}$ & $2.80\pm 0.12$ & --\\
$S_{8.0\mu\rm m}$ & $4.2\pm 0.2$ & --\\
$S_{24\mu\rm m}$ & $12.1\pm 0.3$ & --\\
$S_{12\mu\rm m}$ & $< 23$ & $< 38$ \\
$S_{25\mu\rm m}$ & $< 42$ & $< 58$ \\
$S_{60\mu\rm m}$ & $< 40$ & $< 38$ \\
$S_{100\mu\rm m}$ &$< 84$ & $< 145$ \\
\hline
\end{tabular}
\end{center}
\end{table}
%__________________________________________________________________

\begin{table}
\caption{SED parameters for the two blazars. Peak fluxes, $\nu F(\nu)$, in $\hbox{erg}\,\hbox{cm}^{-2}\,\hbox{s}^{-1}$.} \begin{center}\label{tab:params} \begin{tabular}{lcc}\hline Name &  [HB89]0906+015 & PKS0907+022 \\ \hline $\log_{10}(\nu_p^{s})$[Hz] & 12.6 & 13.3 \\ $\log_{10}(\nu_p^{s}F(\nu_p^{s}))$ & -11.5 & -12.0 \\ $\log_{10}(\nu_p^{IC})$[Hz] & 20.9 & --\\ $\log_{10}(\nu_p^{IC}F(\nu_p^{IC}))$ & -10.3 & -- \\ \hline
\end{tabular} \end{center} \end{table}

%______________________________________________
\section{Spectral energy distributions of [HB89] 0906+015 and PKS 0907+022}\label{sec:SED}

The photometric data for the two previously known blazars are shown in Figs.~\ref{fig:sed1} and \ref{fig:sed2}. To the data available from NED, we added the H-ATLAS fluxes, the UKIDSS data (Lawrence et al. 2007), new ATCA measurements around 20 GHz, new GMRT measurements at 332 MHz (Jarvis et al., in preparation), new \textit{Spitzer} data (Jarvis et al., in preparation) for [HB89] 0906+015 along with $3\sigma$ upper limits that we obtained from IRAS maps (see Table~\ref{tab:HA_fluxes}). The ATCA measurements were performed by M. Massardi and L. Bonavera, in the framework of the Planck-ATCA Coeval Observation (PACO) project.

The blazar [HB89] 0906+015 was detected by the WMAP satellite (source 215 in the NEWPS5yr catalog; Massardi et al. 2009) and also by the Fermi/Large Area Telescope (LAT) (Abdo et al. 2010; note that in the previous bright gamma-Ray source list the $\gamma$-ray source was identified with PKS 0907+022). The shape of its SED exhibits the typical double hump distribution, attributable to synchrotron and inverse Compton, respectively.

To determine the SEDs, we followed the approach of Abdo et al. (2009). First we fitted the part of the SEDs dominated by the synchrotron emission using a third degree polynomial. From these fits, we obtained an estimate of the synchrotron peak energy, $\nu_p^{s}$, and peak intensity, $\nu_p^{s}F(\nu_p^{s})$. Since [HB89] 0906+015 has strong broad emission lines, its optical-UV flux is most likely dominated by the thermal emission produced by the accretion disk, with only a minor contribution from the beamed non-thermal continuum (Ghisellini et al. 2010b). There are indeed clear indications of an optical-UV bump. To avoid biasing estimates of the synchrotron peak frequency and intensity, we excluded these data from the fit. 
Abdo  et al. (2009) demonstrated that the peak of the inverse Compton SED component, $\nu_p^{IC}$, is strongly correlated with the $\gamma$-ray photon index ($\Gamma$). We used their best fit $\nu_p^{IC}$--$\Gamma$ relationship (their Eq. (5)) to estimate $\nu_p^{IC}$ for the HB89 blazar ($\Gamma$=2.74). Finally we obtained the peak intensity by fitting the X-ray to $\gamma$-ray data points with a parabola peaking at $\nu_p^{IC}$ in the $\log(\nu F_\nu)$--$\log \nu$ plane.

The SED parameters are summarized in Table \ref{tab:params}. For PKS 0907+022, since no X-ray nor $\gamma$-ray measurements are available, we could only use the data fitting method for the synchrotron peak. In both cases, the available data from different instruments are non-simultaneous. As for all blazars, they may therefore by affected by variability, which causes additional uncertainty in the derived spectral parameters. This adds to the uncertainties in estimates of spectral parameters. Based on the best-fit model values of $\nu_p^s$, both blazars can be classified as LSPs; however, an ISP classification cannot be ruled out.

%__________________________________________________________________

\section{Conclusions}\label{sec:concl}

The \textit{Herschel} detection of two blazars in the SDP area is interesting in a number of ways. First of all, it confirms that the H-ATLAS survey can play an important role in determining properties of blazars in a crucially important spectral region.
With an area of $550\,\hbox{deg}^2$, the full H-ATLAS should yield a sample of $\simeq 80$ blazars, the first statistically significant sample selected at sub-mm wavelengths. The large area is of crucial importance in this context, because the blazar counts are expected to be relatively flat. The H-ATLAS survey will extend by about one order of magnitude downwards in flux the blazar number counts from the {\sc Planck} survey, which is predicted to have a (95\% reliability) detection limit of 430 mJy at 545 GHz (Leach et al. 2008).

The number of detected blazars is consistent with the predictions of the De Zotti et al. (2005) model which yields $\simeq 0.15\,\hbox{blazars\ deg}^{-2}$ brighter than $50\,$mJy at $500\,\mu$m, i.e., 2.4 objects in the 16 $\hbox{deg}^{-2}$ SDP area. 
An observational estimate of the blazar surface density at 5 GHz was obtained by Padovani et al. (2007; see their Figs. 4 and 6). Taking into account the steepening at frequencies approaching the synchrotron peak, the effective 5 GHz to $500\,\mu$m spectral index yielded by the model is $\simeq 0.2$. The $500\,\mu$m flux limit of 50 mJy then corresponds to 130 mJy at 5 GHz; our results are then in very good agreement with the corresponding blazar surface density found by Padovani et al. (2007) and the relatively flat spectral index assumed by the de Zotti et al. model. If this result is confirmed by data of the full H-ATLAS blazar sample, future models will need to explain in addition evidence of a substantial spectral curvature downwards from cm to mm wavelengths (Waldram et al. 2007; Sadler et al. 2008).

The blazar catalog of Massaro et al. (2009) lists three more blazars in the SDP field: PKS0858-004 with $S_{1.4\rm GHz}=325\,$mJy, 1RXS J085920.6+0047 with $S_{1.4\rm GHz}=39\,$mJy, and 1RXS J085749.8+0135 with $S_{1.4\rm GHz}=89\,$mJy. None of them has a plausible counterpart among the H-ATLAS sources. In the case of PKS0858-004, this implies that the effective 1.4 GHz to $500\,\mu$m spectral index has to be steeper than 0.3, i.e., steeper than the effective indices of the two detected blazars (0.13 and 0.25), but still within the range observed for these sources. 

Based on the best-fit model estimate of the synchrotron peak frequency, $\nu_p^s$, both [HB89] 0906+015 and PKS 0907+022 fall in the low frequency synchrotron peaked (LSP) category. This suggests that the H-ATLAS sample will set strong constraints on the abundance of blazars with low ($\simlt 10^{13}\,$Hz) values of $\nu_p^s$, which, in the blazar sequence scenario, are those with more powerful jets, more luminous accretion disks and higher black hole masses (Ghisellini et al. 2010a).

\begin{acknowledgements}
We are grateful to the referee, P. Padovani, for very useful comments. Thanks are due to the PACO collaboration for having made available the data on the 2 blazars. MM and LB thanks the staff at the Australia Telescope Compact Array site, Narrabri (NSW), for the valuable support they provide in running the telescope. Work partially supported by the Italian Space Agency (contract I/016/07/0
``COFIS'' and ASI/INAF Agreement I/072/09/0 for the Planck LFI Activity of Phase E2). This research has made use of the NASA/IPAC Extragalactic Database (NED) which is operated by the Jet Propulsion Laboratory, California Institute of Technology, under contract with the National Aeronautics and Space Administration.

\end{acknowledgements}

\end{document}